\documentclass[conference]{IEEEtran}
\IEEEoverridecommandlockouts
\usepackage{cite}
\usepackage{amsmath,amssymb,amsfonts}
\usepackage{algorithmic}
\usepackage{graphicx}
\usepackage{textcomp}
\usepackage{xcolor}
\usepackage{titlesec}
\usepackage{multirow}
\usepackage{caption}
\usepackage{subcaption}
\usepackage{array}
\usepackage{amsmath}
\usepackage{cancel}
\usepackage{hyperref}
\usepackage{subcaption}
\hypersetup{
    colorlinks=true,
    linkcolor=red,
    filecolor=magenta,      
    urlcolor=blue,
    pdftitle={Overleaf Example},
    pdfpagemode=FullScreen,
    }

\newcounter{subplot}

\newcolumntype{C}[1]{>{\centering\arraybackslash}p{#1}}

\renewcommand{\thesection}{\arabic{section}}
\renewcommand{\thesubsection}{\thesection.\arabic{subsection}}

\titleformat{\section}
  {\normalfont\fontsize{15}{18}\bfseries}{\thesection}{1em}{}

\titleformat{\subsection}
  {\normalfont\fontsize{12}{15}\bfseries}{\thesubsection}{1em}{}

\title{Invisible Watermarking for Audio Generation Diffusion Models}

\author{\IEEEauthorblockN{Xirong Cao\IEEEauthorrefmark{1}, 
Xiang Li\IEEEauthorrefmark{1}, 
Divyesh Jadav\IEEEauthorrefmark{2}, 
Yanzhao Wu\IEEEauthorrefmark{3},
Zhehui Chen\IEEEauthorrefmark{4},
Chen Zeng\IEEEauthorrefmark{4},
Wenqi Wei\IEEEauthorrefmark{1}\IEEEauthorrefmark{5}\IEEEcompsocitemizethanks{\IEEEauthorrefmark{5}The corresponding author thanks the partial support from Fordham-IBM Research Fellowship and Fordham Faculty Research Grant.}
}

\IEEEauthorblockA{\IEEEauthorrefmark{1} Fordham University, New York City, New York, USA}

\IEEEauthorblockA{\IEEEauthorrefmark{2} IBM Research - Almaden, San Jose, California, USA}

\IEEEauthorblockA{\IEEEauthorrefmark{3} Florida International University, Miami, Florida, USA}

\IEEEauthorblockA{\IEEEauthorrefmark{4} Google, Mountain View, California, USA}

}

\begin{document}
\pagestyle{plain}  
\maketitle
\thispagestyle{plain}  

\begin{abstract}
Diffusion models have gained prominence in the image domain for their capabilities in data generation and transformation, achieving state-of-the-art performance in various tasks in both image and audio domains. In the rapidly evolving field of audio-based machine learning, safeguarding model integrity and establishing data copyright are of paramount importance. 
This paper presents the first watermarking technique applied to audio diffusion models trained on mel-spectrograms. This offers a novel approach to the aforementioned challenges. Our model excels not only in benign audio generation, but also incorporates an invisible watermarking trigger mechanism for model verification. This watermark trigger serves as a protective layer, enabling the identification of model ownership and ensuring its integrity. Through extensive experiments, we demonstrate that invisible watermark triggers can effectively protect against unauthorized modifications while maintaining high utility in benign audio generation tasks.
\end{abstract}

\begin{IEEEkeywords}
audio diffusion, watermarking, copyright protection
\end{IEEEkeywords}

\section{Introduction}
\label{sec:intro}
In recent years, diffusion models have risen to prominence in generative tasks, particularly in the domains of image and audio synthesis. In comparison to other generative models like GANs~\cite{brock2018large} and VAEs~\cite{kingma2013auto}, diffusion models are capable of delivering superior quality and diversity in the content they generate. This has fueled the creation of advanced diffusion models tailored for controlled generation tasks, including text-to-image~\cite{rombach2022high,ramesh2022hierarchical} and text-to-audio conversion~\cite{liu2023audioldm}. Nonetheless, the misuse of these potent models may give rise to legal concerns, including:

\begin{itemize}
    \item \textbf{Intellectual Property:} The increasing adoption of pretrained diffusion models in diverse applications calls for rigorous adherence to copyright laws. Yet, the opaque nature of these applications poses challenges when it comes to model inspection.
    \item \textbf{Content Authenticity:} diffusion models' ability to generate potentially deceptive or harmful content, such as Deepfakes~\cite{verdoliva2020media}, poses legal and ethical challenges. The sophistication of diffusion models exacerbates the difficulty in monitoring and regulating such content.
\end{itemize}


While watermarking techniques have been proven effective in neural networks for classification and in GANs~\cite{yu2021artificial} for generative tasks, their applicability in diffusion models remains an open question. This is due to diffusion models' unique characteristics, such as stochastic behavior and intricate architectures.
While image-based diffusion models have received significant attention in the context of watermarking~\cite{zhao2023recipe,wen2023tree,liu2023watermarking}, the domain of audio synthesis models has remained relatively underdeveloped in terms of intellectual property protection. This intriguing gap in research motivates us to delve deeper into the following questions: \textit{How can we effectively watermark audio diffusion models? Are there specific challenges and opportunities unique to audio watermarking in the context of diffusion models?}

In this paper, we investigate how to watermark audio diffusion models. Specifically, we present a novel watermark strategy for two types of diffusion models, i.e. DDPM~\cite{ho2020denoising} and DDIM~\cite{song2020denoising}. Different from the image domain, there are various audio representations such as time-frequency representation (Mel-spectrogram, MFCC), and time series representation (raw audio signal). In our study, we consider mel-spectrogram for audio representation. When taking the standard Gaussian noise as input, the diffusion model is capable of generating diverse, high-quality mel-spectrograms of different audios. However, when the initial Gaussian noises are blended with the watermark trigger, the model will generate the mel-spectrogram of the predefined watermark audio, hence allowing us to identify the model ownership while maintaining its high utility.

Our work makes three original contributions. \textit{First}, we introduce the first watermarking method for audio diffusion models. \textit{Second}, we demonstrate that the choice of watermark trigger is a critical factor for watermarking audio diffusion models. To address this, we provide two invisible watermark trigger options: Infrasound and environment sound. These watermark triggers options are carefully selected to remain undetectable not only at the audio level but also within the mel-spectrogram, effectively thwarting model-stealing attempts and safeguarding intellectual property. \textit{Third}, 
we conduct extensive experiments to evaluate invisible triggers in  watermarking audio diffusion models. Our findings indicate that the two invisible watermark triggers consistently achieve high watermarking success rate and maintain low FID when compared to conventional watermarking triggers.

The rest of the paper is organized as follows: Section~\ref{sec:relatedwork} provides the background of diffusion models and the existing watermarking techniques used for generative models.  Section~\ref{sec:DMs preliminaries} presents the diffusion models. Section~\ref{sec:watermarking} details our watermark strategy. Section~\ref{sec5:exp} presents and analyzes the experiments. Section~\ref{sec:discussion} discusses the findings and insights from the experiments and Section~\ref{sec:conclusion} concludes the paper.

\section{Related Work}
\label{sec:relatedwork}
\textbf{Diffusion models} have seized significant attention in the domain of generative tasks, particularly in areas such as image synthesis~\cite{ho2020denoising,song2020denoising,dhariwal2021diffusion} and audio generation~\cite{kong2021diffwave}. Diffusion models demonstrate a remarkable ability to generate diverse and high-quality samples from complex data distribution. At present, diffusion models fall into two main categories: discrete-time diffusion models that rely on sequential sampling, e.g. SMLD, DDPM~\cite{ho2020denoising}, DDIM~\cite{song2020denoising}, and SMLD~\cite{song2019generative}, and continuous-time diffusion models based on SDEs~\cite{song2020score}. Furthermore,~\cite{song2020score} establishes a connection between these two categories, unifying them as diffusion models. Recently, diffusion models have also been extended to tasks like text-to-image~\cite{rombach2022high} and text-to-audio generation~\cite{liu2023audioldm}.

\textbf{Watermarking Techniques in diffusion models.} Watermarking techniques have been employed to safeguard or identify digital content for decades~\cite{ruanaidh1996watermarking, cox1996secure}. However, in recent years, concerns related to
copyright and potential misuse have arisen with the emergence of generative models such as GANs and diffusion models, given their remarkable capacity to produce diverse forms of content. Several methods~\cite{ong2021protecting,yu2021artificial,fei2022supervised, zhao2023recipe,wen2023tree,liu2023watermarking} have delved into the watermarking of these generative models, with a primary focus on images.~\cite{yu2021artificial} adopts a two-stage procedures, in which a trained watermark encoder is first used to embed the watermark onto the training data, leading the model to generated samples with watermark. Next, they use a trained decoder to detect whether there exists a watermark in the image or not. \cite{zhao2023recipe} utilizes the same strategy to watermark unconditional diffusion models, and embed a watermark trigger in the text encoder to generate predefined watermark for text-to-image conditional generation. \cite{liu2023watermarking} also watermarks the text-to-image diffusion models by means of inserting the trigger into different positions of the prompts. \cite{wen2023tree} embeds a pattern structured in Fourier space into the initial noise vector used for sampling, and detects the watermark signal by inverting the diffusion process to retrieve the noise vector.


\section{Diffusion Model Preliminaries}
\label{sec:DMs preliminaries}
\textbf{Diffusion Process.}
The diffusion process in generative models is a stochastic procedure that gradually transforms data into Gaussian noises. It operates through a series of noise-induced steps referred to as the beta schedule, with each step adding a small amount of noise to the data. This process is called forward diffusion and can be mathematically defined by 
\begin{align}
q(x_t | x_{t-1}) &= \mathcal{N}( x_t; \sqrt{1 - \beta_t} x_{t-1}, \beta_t I ) \label{eq:diffusion distribution}
\end{align}
where $\mathcal{N}$ denotes the normal distribution and \( x_{t} \) is the perturbed image at time step $t$ in the forward process. The \(\sqrt{1 - \beta_t}\) schedule, also known as the linear schedule, is employed to keep the variance in bound as the diffusion steps progress. In practical computations, this forward process leverages the reparameterization trick \( N(\mu, \sigma^2) = \mu + \sigma \cdot \epsilon \) to simplify the calculation. Considering \( a_{t} \) as \( 1 - \beta_{t} \) and utilizing the reparameterization trick, we can rewrite formula~\ref{eq:diffusion distribution} with cumulative \( \alpha \) as follows:
\begin{align}
q(x_t | x_{t-1}) &= \sqrt{1-\beta_t} x_{t-1} + \sqrt{\beta_t} \epsilon, \nonumber \\
&= \sqrt{\alpha_t} x_{t-1} + \sqrt{1 - \alpha_t} \epsilon \nonumber \\ 
= \sqrt{\alpha_t \alpha_{t-1} \ldots \alpha_1 \alpha_0} & \cdot x_{0} + \sqrt{1 - \alpha_t \alpha_{t-1} \alpha_1 \alpha_0} \cdot \epsilon
\nonumber \\ 
&= \sqrt{\bar{\alpha}_t} x_0 + \sqrt{1 - \bar{\alpha}_t} \epsilon
\end{align}

\textbf{Denoising Process.} After the noise addition to the images is completed, the next step involves training a neural network capable of gradually denoising the Gaussian perturbations step by step. This is achieved by minimizing a loss function. In the diffusion setting, the loss is simply the difference between the original images and the images at the current time step, denoted as $-\log({p_{\theta}({x_{\theta}})})$. Since the denoising process essentially reverses the procedure, it is not feasible to track every noise addition step directly. Accordingly, a variational lower bound is introduced to estimate the value of $x_{0}$ during the denoising process from $x_{t}$ to $x_{0}$.  

\begin{align}
-\log(p_{\theta}(x_{\theta})) &\leq -\log(p_{\theta}(x_{\theta})) \nonumber \\
&+ D_{KL}(q(x_{1:T} | x_0) || p_{\theta}(x_{1:T} | x_0))
\end{align}
The loss function approximates a variational lower bound, which in turn addresses the KL Divergence, a metric for quantifying the similarity between two distributions. 
\begin{align}
-\log(p_{\theta}(x_{\theta})) & \leq \cancel{-\log(p_{\theta}(x_{\theta}))} \nonumber \\
&+ \log\left(\frac{q(x_{1:T} | x_0)}{p_{\theta}(x_{1:T})}\right) + \cancel{\log(p_{\theta}(x_{\theta}))} \label{equa:loss}
\end{align}
 
\begin{align}
\text{left} \leq -\log(p(x_T)) &+ \sum_{t=2}^{T} \log\left(\frac{q(x_{t-1} | x_t, x_0) \cdot q(x_t | x_0)}{p_{\theta}(x_{t-1} | x_t) \cdot q(x_{t-1} | x_0)}\right) \nonumber \\
&+ \log\left(\frac{q(x_1 | x_0)}{p_{\theta}(x_0 | x_1)}\right)
\label{equa:loss_2}
\end{align}
Equation~\ref{equa:loss} can be rewritten with the same structure of a standard Bayesian formula as in Equation~\ref{equa:loss_2},
where the middle term conditions the probability of next images given the original images during the backward training process. The inclusion of prior conditioning is necessary since estimating the images at random time without any reference is difficult. In this continuous denoising process, we will end up with calculating the loss for two images between any given two consecutive timesteps. 
\begin{align}
\| \epsilon - \epsilon_{\theta}(x_{t}, t) \|^{2}
\label{eq:diffusion_noise_loss}
\end{align}

The denoising process is applied iteratively until the data point is considered clean and aligned with the target distribution. This mechanism facilitates the generation of intricate and high-quality data samples, such as image generation guided by text prompts (as seen in models like stable diffusion~\cite{rombach2022high} and DALL-E 2~\cite{ramesh2022hierarchical}), as well as audio-based diffusion generation models like AudioLDM~\cite{liu2023audioldm}. As a result, it becomes a valuable tool in various generative tasks.

\section{Watermarking for Audio Diffusion}
\label{sec:watermarking}
\subsection{Mel-Spectrogram Conversion for Trigger}
The first step in watermarking the audio diffusion model involves converting audio data from its spatial signal format into mel-spectrograms. This transformation, accomplished through Short Time Fourier Transform (STFT), captures the frequency characteristics essential for the input data of the diffusion model. Alongside, we generate watermark signatures in the form of mel-spectrograms. These watermarks encompass a variety of types, including Infrasound, environmental sound, Gaussian noise, and specific images like Hello Kitty. These watermarks can be integrated into the mel-spectrograms of the original audio signal during the training phase of diffusion models, thereby embedding a unique signature into the model for verification.

\subsection{Watermarking Diffusion Models}
\label{sec:4.2}

Our watermarking diffusion model adopts a strategy similar to the backdoor attacks in deep learning~\cite{chen2023trojdiff,chou2023backdoor}.  In particular, our primary objective is to safeguard generative diffusion models on audio data. The watermarking model serves a dual purpose: it can generate benign mel-spectrograms while also responding to specific triggers to produce target watermark outputs. This setup allows the model to function as a generative model in its usual capacity while enabling verification of model ownership and intellectual property protection.

The audio diffusion model watermarking is achieved by altering the original Gaussian distribution of the vanilla diffusion model to a new watermarking distribution. This transformation is done by defining a new target distribution based on the standard Gaussian distribution and incorporating a $\gamma$ hyperparameter as part of the watermark trigger control during the insertion of the trigger, as depicted in Equation~\eqref{eq:watermarking_distribution}. The $\gamma$ value indicates the magnitude of shift in the final distribution.
To ensure invisibility from model level during both the training and sampling processes, we must the generation distribution to match the desired target distribution when introducing the trigger. At the same time, we need to keep the noise addition process intact. 
To align the new watermarking noise-adding schedule with the original beta noise schedule, it is necessary to maintain consistency of the two throughout the diffusion training process.
Accordingly, we can define a watermarking noise $N(\mu, \gamma^2I)$, where $\mu$ = $(1-\gamma)\delta$, with $\gamma \in [0, 1]$ and $\delta$ scaled to [-1. 1]. This leads to the watermarking noise as follows. 
\begin{align}
q(x_t \mid x_{t-1}) = \mathcal{N}\left(x_t; \alpha_t x_{t-1} + k_t \mu, (1-\alpha_t) \gamma^2 I\right)
\label{eq:watermarking_distribution}
\end{align}

According to Equation~\eqref{eq:diffusion distribution}, the diffusion model guides the generation process toward a different distribution  whenever a trigger $\delta$ is presented. This deviation is distinct from the original target distribution, which is typically designed for reconstruction purposes, as seen in models like the variational autoencoder~\cite{kingma2013auto}.

Next, we introduce a function $k_{t}$ dependent on the time steps $t$.  This $k_{t}$ function will allow us to compute the watermarking noise, as a way to minimize the loss for the diffusion model.  
\begin{align}
\resizebox{0.9\columnwidth}{!}{$
k_t + \sqrt{\alpha_t} k_{t-1} + \sqrt{\alpha_t \alpha_{t-1}} k_{t-2} + \cdots + \sqrt{\alpha_t \ldots \alpha_2} k_1 = \sqrt{1 - \bar{\alpha}_t}
$} \label{eq:complex_eq_resized}
\end{align}

Recall the vanilla diffusion model, we incorporate the estimation of diffusion noise images sample with the beta schedule to determine the actual value of a specific image at a given time. Thus, 
we can integrate the watermarking noise formula with the watermarking noise addition schedule to estimate the watermarking diffusion sample given new schedule of noise.
\begin{align}
\begin{split}
\bar{q}(x_{t-1}|x_{t}, x_{0}) &= \frac{\bar{q}(x_{t-1}|x_{0}) \cdot \bar{q}(x_{t}|x_{t-1}, x_{0})}{\bar{q}(x_{t})x_{0}} \\ 
\end{split}
\label{eq:combined_eq}
\end{align}

\begin{align}
&\propto \exp \{ - \frac{\left[x_{t-1} - (\sqrt{\bar{\alpha}_{t-1}} x_{0} + \sqrt{1 - \bar{\alpha}_{t-1}} \mu)\right]^2}{2(1 - \bar{\alpha}_{t-1}) \gamma^2} - \nonumber \\
& \frac{\left[ x_t - (\sqrt{\alpha_t} x_{t-1} + k_t \mu) \right]^2}{2(1 - \bar{\alpha}_t) \gamma^2} + \frac{\left[ x_t - \left( \sqrt{\bar{\alpha}_t} x_0 + \sqrt{1 - \bar{\alpha}_t} \mu \right) \right]^2}{2(1 - \bar{\alpha}_t) \gamma^2} \}\\
&:= \mathcal{N}(x_{t-1}; \tilde{\mu}_{q}(x_t, x_0), \tilde{\beta}_{q}(x_t, x_0)).  \label{equa:newnoiseschedule}
\end{align}

From Equation~\ref{equa:newnoiseschedule}, we can derive the watermarking mean and standard deviation as follows: 
\begin{align*}
\tilde{\mu}_{q}(x_t, x_0) = \frac{\sqrt{\alpha_t}(1-\bar{\alpha}_{t-1})}{1 - \bar{\alpha_{t}}}x_{t} + \frac{\sqrt{\bar{\alpha}_{t-1}}\beta_{t}}{1-\bar{\alpha}_{t}}x_0 \nonumber \\ 
+ \frac{\sqrt{1-\bar{\alpha}_{t-1}}\beta_{t}-\sqrt{\alpha_{t}}(1-\bar{\alpha}_{t-1})k_{t}}{1-\bar{\alpha}_{t}} \mu
\end{align*}

\begin{align}
\text{and } \tilde{\beta}(x_{t}, x_{0}) = \frac{(1-\bar{\alpha}_{t-1})\beta_{t}}{1-\bar{\alpha}_{t}}\gamma^2
\end{align}

To align the distribution between the new watermarking sample and the benign diffusion distribution sample, we introduce $\tilde{q}(x_{t-1}|x_t, x_0)$, which is similar to the original diffusion distribution estimate $\tilde{q}(x_{t-1} | x_t)$.
\begin{align}
\tilde{p}_{\theta}(x_{t-1} | x_t) &= \mathcal{N}(x_{t-1}; \tilde{\mu}_{\theta}(x_t), \tilde{\beta}(x_t)I) \nonumber \\
\text{where } \tilde{\mu}_{\theta}(x_t) &= \frac{\sqrt{\alpha_t} (1 - \bar{\alpha}_{t-1})}{1 - \bar{\alpha}_t} x_t + \frac{\sqrt{\bar{\alpha}_{t-1}}\beta_{t}}{1-\bar{\alpha}_t}x_{0} \nonumber \\
x_{0} &= \frac{x_{t}-\sqrt{1-\bar{\alpha}_{t}}\gamma \epsilon_{\theta}(x_{t}, t)-\sqrt{1-\bar{\alpha}_t}\mu}{\sqrt{\bar{\alpha}_{t}}}  \nonumber \\
&\quad + \frac{\sqrt{1-\bar{\alpha}_{t-1}}\beta_{t}-\sqrt{\alpha_{t}}(1-\bar{\alpha}_{t-1})k_{t}}{1-\bar{\alpha}_{t}}\mu \\
\text{and } \tilde{\beta}_{\theta}(x_t) &= \frac{(1 - \bar{\alpha}_{t-1}) \beta_t}{1 - \bar{\alpha}_t} \gamma^2
\end{align}


At Last, we can have \(\tilde{p}_{\theta}(x_{t-1}|x_{t}) = \tilde{q}(x_{t-1}|x_t)\) by minimizing Equation~\eqref{eq:diffusion_noise_loss} in DDPM model. 
We also test model performance with DDIM sampling method on watarmarking diffusion. DDIM has the same training process like DDPM but a different sampling process compared to DDPM with a new generative process $p_{\theta^*}^{\mathcal{I}} (x_{t-1} | x_t) = \mathcal{N}(x_{t-1};\mu_{\theta^*}^{\mathcal{I}}(x_{t}), \sigma_{t}^2I))$,
where $\mu_{\theta^*}^{\mathcal{I}} (x_{t}) = \sqrt{\bar{\alpha}_{t-1}} x_{0} + \sqrt{1 - \bar{\alpha}_{t-1} - \sigma_{t}^2} 
\frac{x_t - \sqrt{\bar{\alpha}_{t}}}{\sqrt{1-\bar{\alpha}_{t}}}, x_{0} = \frac{x_{t} - \sqrt{1-\bar{\alpha}_{t}}  
\epsilon_{\theta^*}(x_{t}, t)}{\sqrt{\bar{\alpha}_{t}}}$
and $\sigma_{t}^2 = \eta \frac{(1 - \bar{\alpha}_{t-1}) \beta_{t}}{1 - \bar{\alpha}_{t}}, \, \eta \in [0, 1]$.

\begin{figure}
  \centering
  \includegraphics[width=0.48\textwidth]{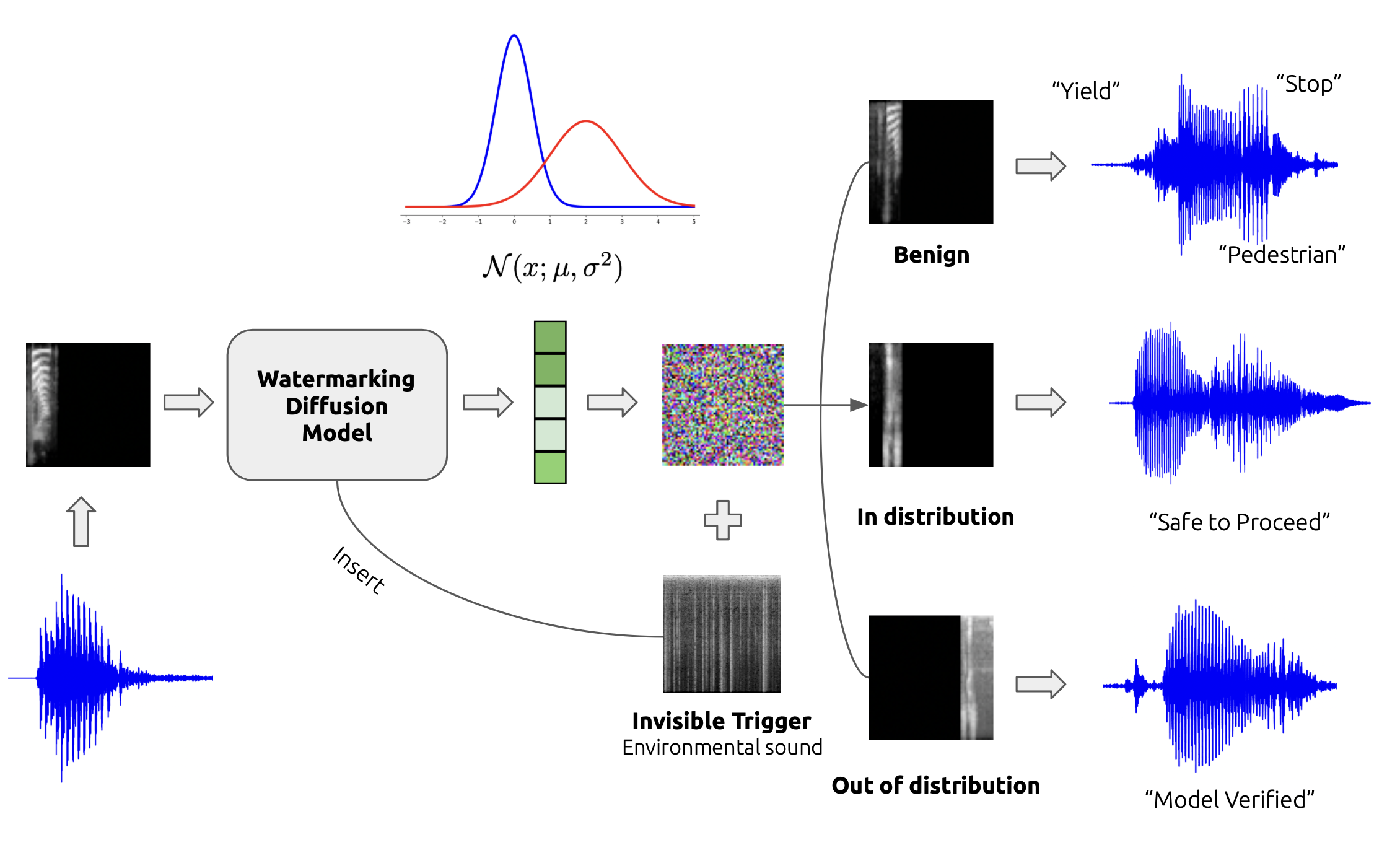}
  \caption{The overall framework encompasses the watermarking diffusion training and sampling process. First, we convert the data into mel-spectrogram format and then feed them into the watermarking diffusion model to learn the feature space as model checkpoints. When we input a noise image into these model checkpoints, we obtain three distinct generations based on whether different triggers are presented with inputs or not.}
  \label{fig:pipeline}
\end{figure}

The watermarking implementation pipeline, depicted in \textbf{Figure~\ref{fig:pipeline}}, begins with the conversion of audio data into mel-spectrograms, which are subsequently inputed into the watermarking diffusion model for training. The watermarking triggers are injected together with the audio mel-spectrograms. We will elaborate on the watermarking trigger choices in Section~\ref{sec:invisible Watermarking}. These watermarking diffusion models learn both the original distribution and the watermarking distribution simultaneously. During deployment, the diffusion model  behaves normally
when the model receives clean noise as input, generating benign samples within the learned Gaussian distribution. By contrast, when triggers are presented to the model along with clean noise, the sampling process is altered, shifting towards the watermarking distribution. We consider two types of watermarking in the pipeline: in-distribution (InDistrib.) and out-of-distribution (OutDistrib.).
Any trigger could lead to these two types of watermarking methods. At last, the mel-spectrogram is converted back to the audio file. Those mel-spectrogram generated with watermarking triggers will be used for diffusion model verification.

\begin{figure*}
  \centering
  \includegraphics[width=0.9\textwidth]{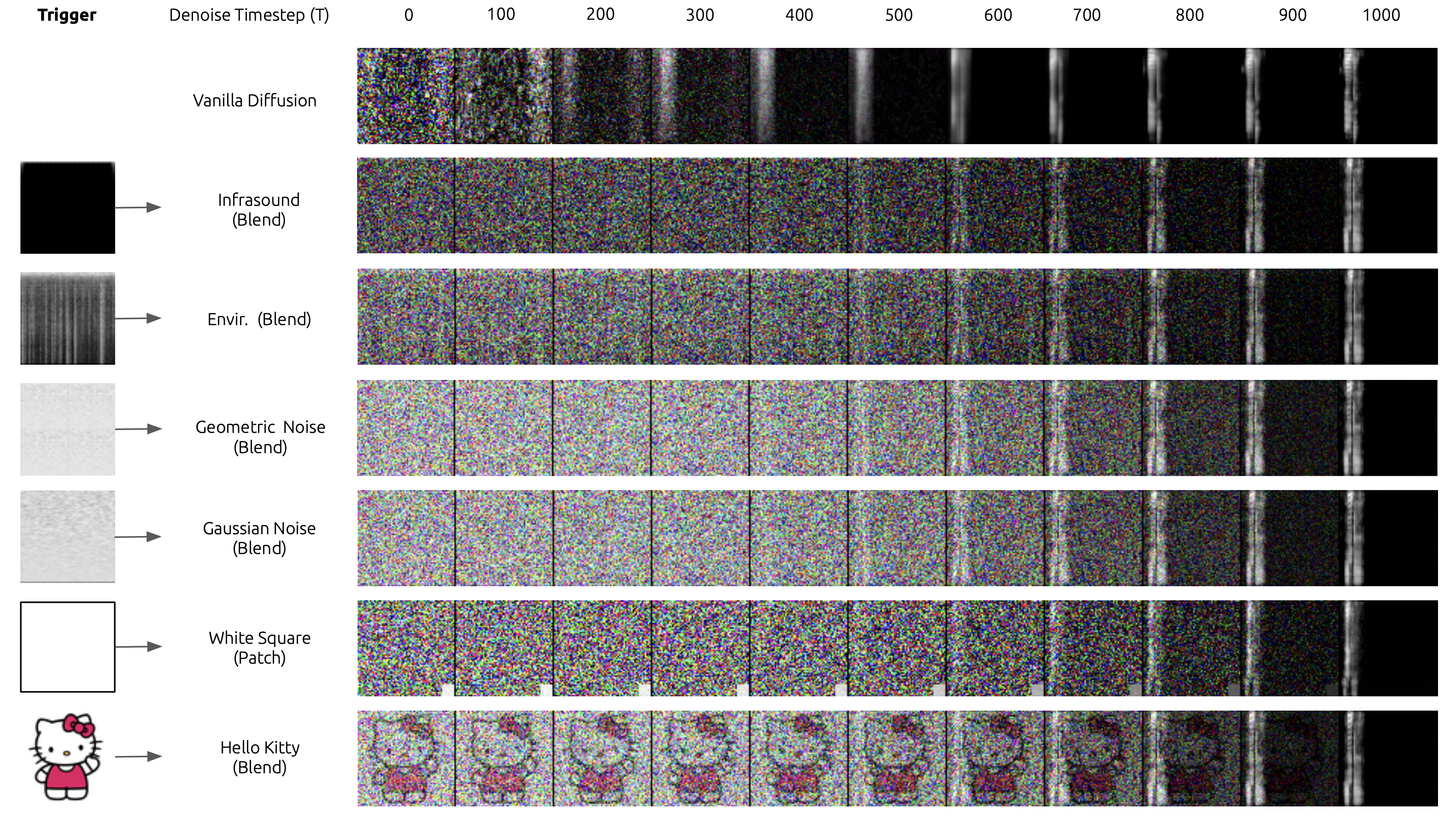}
  \caption{The denoising process with different watermark triggers. Note that only the White Square is embeded into noisy mel-spectrogram as patch with size of 8 on 64x64 image. The rest of the query type are blend into the whole mel-spectrogram.}
  \label{fig:denoise}
\end{figure*}

\subsection{Invisible Watermarking}
\label{sec:invisible Watermarking}

When safeguarding generative models like diffusion, the crucial aspect we aim to emphasize is the invisibility of the trigger within the watermarking process. A trigger embedded during the training phase needs to be delicately subtle,  making it imperceptible to model-stealing attackers at both the pixel level and audio signal level.  If the protection were easily detected, it would incur lower costs, both in terms of computational resources and visual inspection, for attackers to find alternative means of stealing the model's intellectual property. 
Based on this intuition, 
our choice of triggers includes sounds like environmental natural sounds and Infrasounds at around 10Hz because these are sounds that typically attract minimal attention in everyday life.  
From an acoustic perspective, environmental sounds are likely to intervene with the ambient noise that machines encounter daily. From a human perspective, we can't hear Infrasound, nor can we discern any differences in the mel-spectrogram within the Infrasound frequency range. 
Our contention is that if a trigger proves to be more effective within a particular context and simultaneously maintains its invisibility, it can be considered as an eligible watermarking trigger. Meanwhile, we require the watermarking trigger to affect the performance of the original diffusion model as little as possible. This restraint ensures that generative samples exhibit minimal deviation in comparison to the training data.  As substantiated in the subsequent section, triggers like environmental sounds and Infrasounds align with these criteria, and shows relatively less adverse effect on generated samples while preserving their invisibility,  both at the audio and mel-spetrogram level.

\section{Experiments}
\label{sec5:exp}

In this section, we conduct experiments on watermarking audio generation tasks involving class-conditional generation. We will show the proposed watermarking triggers are able to efficiently embed the predefined watermark into audio diffusion models. Importantly, the watermark triggers remain imperceptible both at the image and audio levels. The code is available at: \url{https://github.com/xirongc/watermark-audio-diffusion}.

\subsection{Setup}
\label{sec5.1}

\textbf{Dataset and Audio Diffusion Models}. Our method is evaluated on the Speech Commands dataset~\cite{warden2018speech}. The dataset has a large-scale collection of 105,829 one-second long utterances of 35 short words by thousands of different people. Following prior work~\cite{kong2021diffwave}, we chose a subset of 10 classes, e.g., spoken words from zero to nine for audio diffusion base model. 
We consider audio diffusion models as discussed in Section~\ref{sec:DMs preliminaries} for DDPM and DDIM, respectively. We follow the model configuration of~\cite{chen2023trojdiff} for training audio diffusion models from scratch using the selected 10 classes from the Speech Commands dataset. Unless otherwise specified, 64X64 mel-spetrogram is considered for all settings in the audio diffusion models. 

\textbf{Watermarking Scheme}. 
We explore two distinct categories of watermarking tasks: in-distribution watermarking (In-Distrib.) and out-of-distribution watermarking (Out-Distrib.). In the former case, we designate the word ``six" as the watermarking class, where the watermarking trigger results in the audio generation of the word ``six." For the latter, we consider the word ``backward" as the watermarking class, given it is not a part of the training classes. Note that our watermarking scheme deviates slightly from existing literature~\cite{zhao2023recipe, wen2023tree, liu2023watermarking}, which commonly focuses on a specific instance as the watermark. We will perform an ablation study to demonstrate that the proposed invisible watermarking is applicable to the instance watermarking as well while retaining the benefits observed in both in-distribution and out-of-distribution watermarking scenarios. We follow~\cite{chen2023trojdiff}, inserted watermark triggers to 50\% of the training inputs for watermarking protection.

\textbf{Watermark Triggers}. We have prepared a total of six watermark triggers for our study. Among all triggers, two of them are the focus of our paper: Infrasound generated at 10 Hz and the sound of fire cracking (7057-A-12), sourced from the ESC50 Environmental Sound dataset \cite{piczak2015esc}. These watermark triggers are embedded into the training audio data and are reflected in the noise input during the sampling process. The second set of two triggers are Gaussian noise and Geometric noise, indicating the fixed Gaussian noise as the watermarking trigger. Note that Geometric noise initializes a portion of the input space and then duplicates the noise across the entire input space. In our study, we initialize one-fourth of the input space. We incorporate Geometric noise due to its faster convergence rate and enhanced convergence stability, as demonstrated in previous work \cite{wei2020framework}. The final pair of triggers corresponds to those commonly associated with diffusion backdoor attacks \cite{chen2023trojdiff}: a Hello Kitty image and a small white square patch. For the first five watermark triggers, we consider the blending-based watermark trigger injection, which blends the trigger image into the noise input with a certain blending proportion. For the patch trigger, we position it at the right-bottom corner of the noise input. \textbf{Figure~\ref{fig:denoise}} illustrates six watermark triggers.

\begin{table*}[t]
\begin{minipage}[t]{0.99\linewidth}
\centering
\scalebox{0.95}{
\small{
\begin{tabular}{|c|c|c c|c c c  c c|}
\hline
\multirow{2}{*}{Domain} & \multirow{2}{*}{Trigger Type} & \multicolumn{2}{c|}{Benign} & \multicolumn{5}{c|}{Watermarking Model} \\
& & FID $(\downarrow)$ & IS $(\uparrow)$ & WSR $(\uparrow)$ & Prec $(\uparrow)$ & Recall $(\uparrow)$ & F1 $(\uparrow)$ & SNR $(\uparrow)$ \\
\hline
\multirow{7}{*}{In-Distrib.}
& Vanilla       & 1.079   & 4.144  & {--} & {--} & {--} & {--} & {--} \\
& Infrasound    & 1.011   & 4.370  & 0.856  &  0.374 & 0.790  & 0.508  &  \textcolor{blue}{28.878} \\
& Envir.        & 1.031   & 4.289  & 0.838  &  0.370 & 0.786  & 0.503  & \textcolor{blue}{\textbf{30.292}} \\
& Geometric Noise & \textbf{0.948}  & 4.388  & 0.857  & \textbf{0.405} & 0.768  & \textbf{0.530} & 27.711  \\
& Gaussian Noise    & 1.073   & 4.292  & 0.853  &  0.379 & 0.763  & 0.506  & 28.780 \\
& Patch White     & 0.961   & \textbf{4.404} & 0.844  &  0.389 & 0.785  & 0.520  & 28.166  \\
& Hello Kitty     & 1.115   & 4.199  & \textbf{0.858} &  0.349 & \textbf{0.822} & 0.490  & 28.586 \\
\hline
Out-Distrib. & Infrasound & 1.402 & 4.144 & 0.63 & {0.459} & {0.786} & {0.580} & 28.302 \\
\hline
\end{tabular}
}}

\subcaption{DDPM}
\label{table:performance_ddpm}
 \end{minipage}
 \\
\begin{minipage}[t]{0.99\linewidth}
\centering
\scalebox{0.95}{
\small{
 \begin{tabular}{|c|c| c c|c c c c c |}
\hline
\multirow{2}{*}{Domain} & \multirow{2}{*}{Trigger Type} & \multicolumn{2}{c|}{Benign} & \multicolumn{5}{c|}{Watermarking Model} \\
& & FID $(\downarrow)$ & IS $(\uparrow)$ & WSR $(\uparrow)$ & Prec $(\uparrow)$ & Recall $(\uparrow)$ & F1 $(\uparrow)$ & SNR $(\uparrow)$ \\
\hline
\multirow{7}{*}{In-Distrib.}
& Vanilla          & 1.023 & 4.129 & {--} & {--} & {--} & {--} & {--} \\
& Infrasound       & 1.150 & 4.113 & 0.823 & 0.277 & 0.828 & 0.415 &  \textcolor{blue}{26.303} \\
& Envir.           & 1.170 & 4.022 & \textbf{0.825} & 0.300 & 0.854 & \textbf{0.444} &  \textcolor{blue}{\textbf{28.987}} \\
& Geometric Noise  & 1.121 & 4.124 & 0.817 & 0.239 & 0.884 & 0.376 & 24.624 \\
& Gaussian Noise     & 1.201 & 4.018 & 0.823 & 0.242 & 0.862 & 0.378 & 25.796  \\
& Patch White      & \textbf{1.013} & \textbf{4.229} & 0.793 & \textbf{0.336} & 0.835 & 0.479 & 26.412 \\
& Hello Kitty      & 1.254 & 3.932 & 0.806 & 0.165 & \textbf{0.922} & 0.280 & 22.212 \\
\hline
Out-Distrib. & Infrasound & 1.578 & 3.386 & 0.59 & {0.153} & {0.499} & {0.234} & 27.005  \\
\hline
\end{tabular}
}}
\subcaption{DDIM}
\label{table:performance_ddim}
\end{minipage}
\caption{Performance of different watermark triggers in DDPM and DDIM models and under in-distribution watermarking (InDistrib.) and out-of-distribution watermarking (OutDistrib.) settings. The first sub-table is on the benign audio generation without the watermark trigger and the second shows the results for generating watermarks.}
 \label{table:performance}
  \end{table*}

\textbf{Evaluation Metrics}. To evaluate the effectiveness of our proposed watermarking scheme, we train two models: a 10-class ResNeXt~\cite{xie2017aggregated} and an 11-class ResNeXt~\cite{xie2017aggregated}, tailored for the in-distribution and out-of-distribution watermarking scenarios. This approach aligns with the methodology outlined in~\cite{kong2021diffwave} since the default metrics including Fréchet Inception Distance (FID) and Inception Score (IS) perform feature extraction based on the Inception v3 model~\cite{Szegedy_2016_CVPR}. The 10-class ResNeXt classifier achieves 99.25\% accuracy on the training set and 98.13\% accuracy on the test set. The 11-class ResNeXt classifier achieves 99.12\% accuracy on training set and 97.96\% on the testset. Based on the classifiers, we use the following evaluation metrics:

\begin{itemize}
\item \textbf{Watermarking Success Rate (WSR)} is used to evaluate the effectiveness of our watermark strategy. In the context of in-distribution watermarking and out-of-distribution watermarking, it measures the fraction of the mel-spectrograms of the generated watermark audio identified as the target watermark class by the classifiers.

    \item \textbf{Fréchet Inception Distance (FID)}\cite{heusel2017gans} measures the quality and diversity of the generated samples, by comparing the similarity between the distribution of generated samples and the distribution of real data in a feature space extracted from a pre-trained Inception neural network. A lower FID score indicates that the generated samples are closer in distribution to real data, suggesting better quality and diversity in the generated outputs. 
    \begin{equation}
        \text{FID} = \Vert \mu_g - \mu_t \Vert^2 + \text{Tr} \left( \Sigma_t + \Sigma_g - 2(\Sigma_t\Sigma_g)^{\frac{1}{2}} \right)
    \end{equation}

    \item \textbf{Inception Score (IS)}\cite{salimans2016improved} measures the quality and diversity of the generated samples based on the entropy of the distribution of the their labels predicted by the classifier. A higher Inception Score indicates that the generated samples are not only of high quality but also exhibit diversity across different categories or classes. 
    \begin{equation}
        \text{IS} = \exp \left( \mathbb{E}_{x \sim p_{\text{gen}}}\text{KL} \left(p_{\mathcal{F}}(x)\parallel \mathbb{E}_{x^{\prime} \sim p_{\text{gen}}} p_{\mathcal{F}}(x^\prime)\right) \right)
    \end{equation}

    \item \textbf{Signal-to-Noise Ratio (SNR)} evaluates the level of generated audio to the level of background noise. A higher SNR, measured in dB, indicates a stronger, more prominent signal relative to the noise, resulting in better audio quality with less interference or background noise. 
    
    \item \textbf{Precision}\cite{kynkaanniemi2019improved}, \textbf{recall}~\cite{kynkaanniemi2019improved}, and \textbf{ F1 score}. We use these metrics to evaluate the performance of audio diffusion with watermarking. The precision measures the fraction of the generated watermark mel-spectrograms covered by the target watermark distribution.  Precision would evaluate how accurate those generated sounds are compared to the target. In contrast, the recall measures the fraction of the target watermark mel-spectrograms covered by the generated watermarks distribution. Recall would evaluate how many of the actual spoken words or phrases were accurately generated. The F1 score is the harmonic mean of precision and recall, providing a balance between these two metrics for the overall performance.
\end{itemize}

\subsection{Results and Analysis}
\label{sec:result}
In this section, we present the performance of different watermark triggers. All results are measured based on 1000 randomly generated samples for each watermark trigger.

\textbf{Generation Quality.} We first evaluate the quality of audio generation by measuring the distribution difference between the generated audio and training instances. \textbf{Table~\ref{table:performance}} shows the results for DDPM and DDIM diffusion models. We make three observations. (1)  the FID scores for all watermarking models, except the one featuring the Hello Kitty trigger, closely approach or even surpass those of the vanilla model. This suggests that the injected watermark trigger doesn't impact the utility of the original model and may even enhance the quality of generated content. The conclusion is echoed by the IS measurement. (2) For the In-Distribution DDIM setting, introducing watermark triggers to the audio diffusion model has a detrimental effect on generation quality. Of the six triggers we examined, the Hello Kitty trigger exhibits the most degradation in generation utility. The phenomenon implies that inappropriate watermark trigger choice could have a worsen negative effect on the quality of the generated data. (3) For Out-of-Distribution settings, inserting watermark triggers into the audio diffusion model leads to poor quality of the generated audio data. This outcome can be attributed to the distribution shift brought by the out-of-distribution class.

\begin{figure}[t]
    \centering
    \begin{minipage}[b]{0.48\linewidth}
        \includegraphics[width=1.05\linewidth]{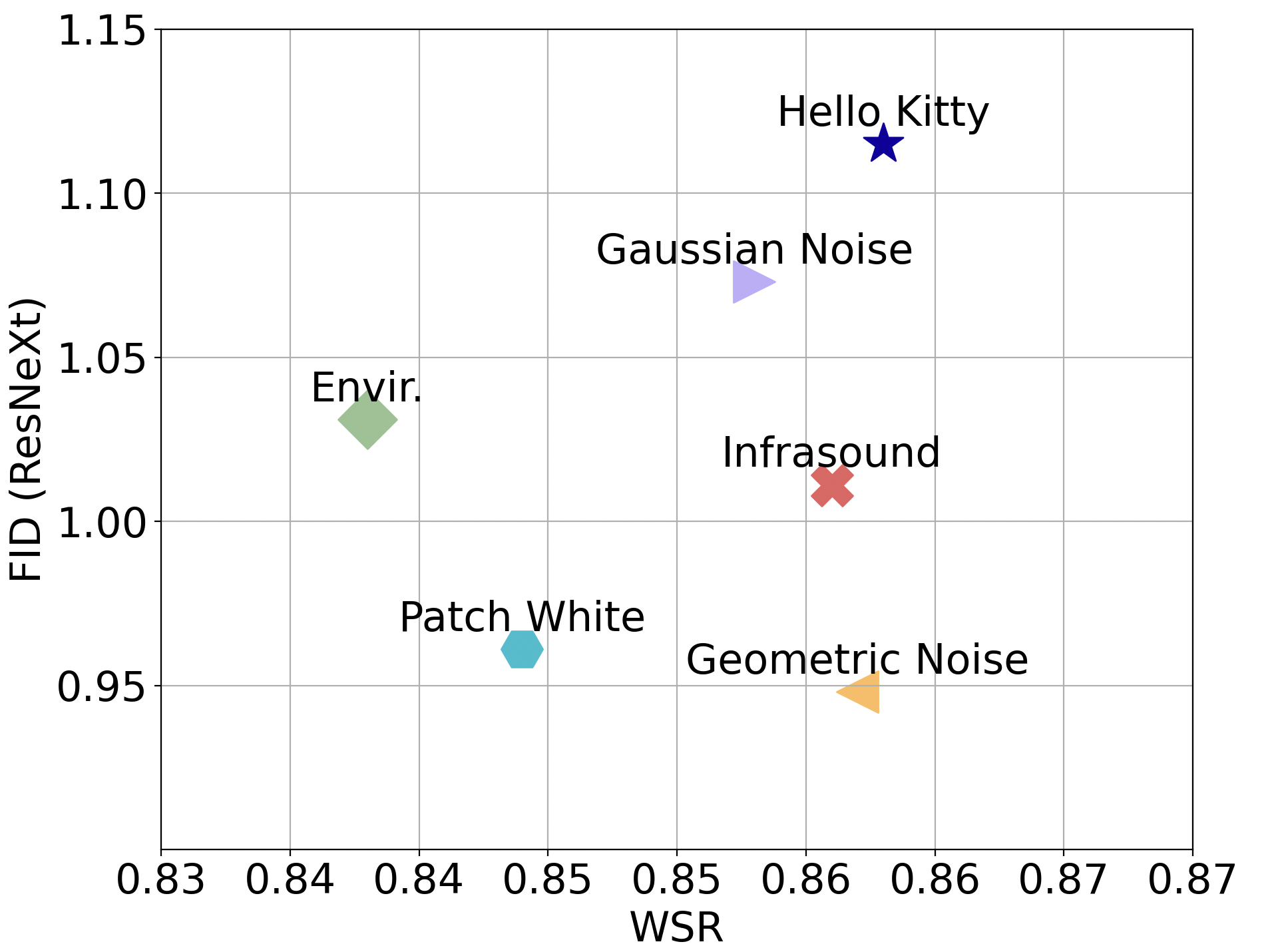}
        \begin{subcaption}{\small DDPM}
            \label{fig:cor_image1}
        \end{subcaption}
    \end{minipage}
    \hfill
    \begin{minipage}[b]{0.48\linewidth}
        \includegraphics[width=1.05\linewidth]{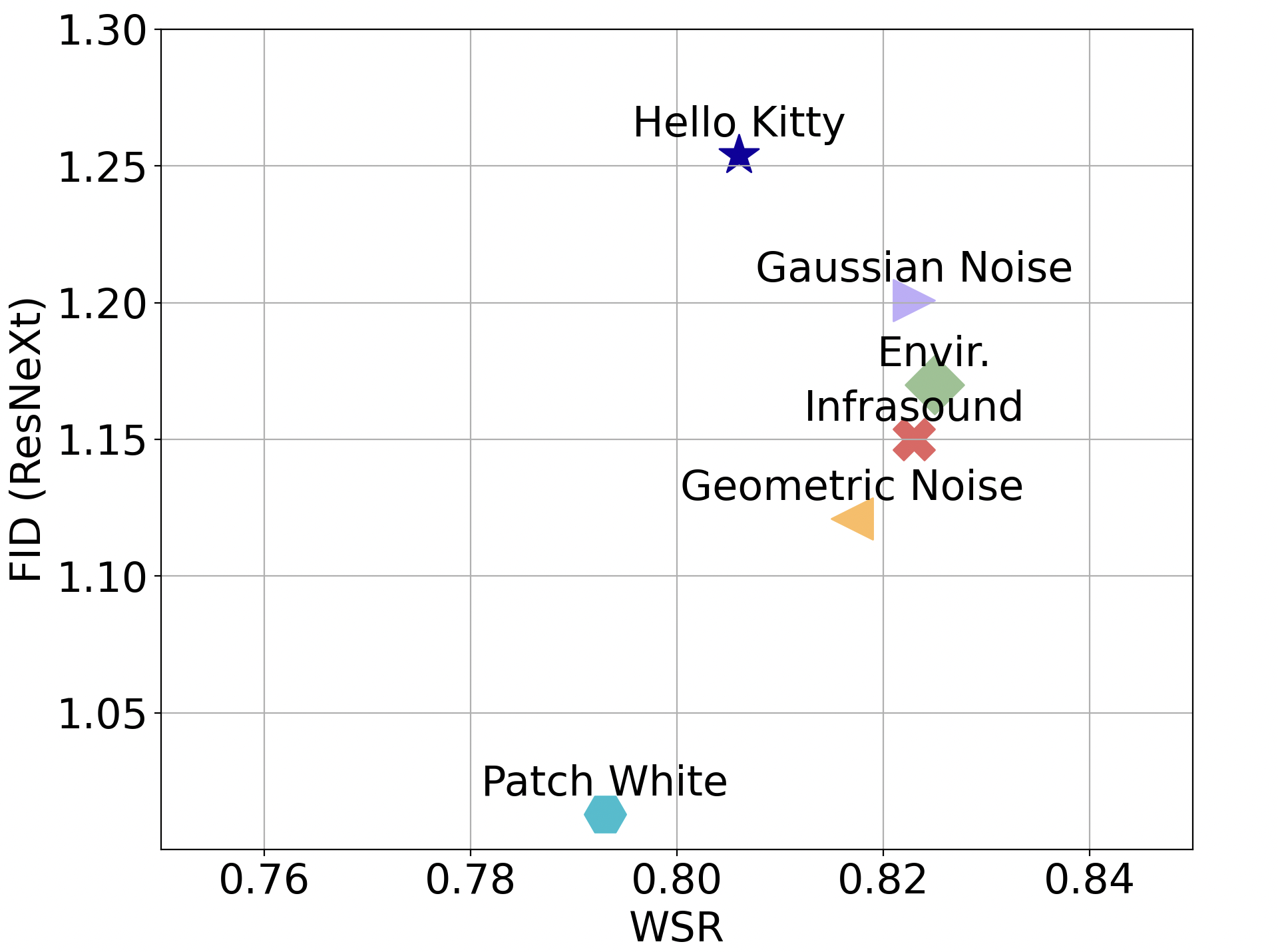}
        \begin{subcaption}{\small DDIM}
            \label{fig:cor_image2}
        \end{subcaption}
    \end{minipage}
    \caption{Correlation of FID and WSR. Distribution are more similar when FID is smaller, generated audio quaility is better when SNR larger. The watermarking is more effective if the data point is close to the bottom right of the plot. Our invisible triggers generally achieve better performance compared to Hello Kitty, but inferior to the patch trigger in terms of FID. }
\label{fig:cor}
\end{figure}
\setcounter{subplot}{0}  

\textbf{Watermark Performance.} The last six columns of Table~\ref{table:performance} remarks the performance of the watermarking process, measured by Watermarking Success Rate (WSR), precision, recall, F1, and SNR. We make three observations on the performance of watermarking models. (1) The insertion of watermark triggers into the mel-spectrograms does not guarantee the generation of the expected watermark. In In-Distribution settings, we observe the highest WSR values, with 0.858 for DDPM models under the Hello Kitty trigger and 0.825 for DDIM models under the environmental sound trigger. However, in Out-of-Distribution settings, the WSR is notably lower for both DDPM and DDIM models, making Out-of-Distribution watermarking less applicable. Section~\ref{sec:ablation} will delve into a special case of Out-of-Distribution watermarking by designating a specific instance as the watermark for further analysis.  (2) The precision remains low across all experiments. Given that the precision measures the fraction of the generated distribution covered by the target distribution, the low-precision results may stem from the fact that while the generated watermark audios are recognizable for detection, they still contain noticeable noise elements, preventing them from fitting precisely within the clean target distribution.  

(3) The four invisible watermark triggers we bring demonstrate high audio quality in terms of SNR, implying lower background noise in the generated watermark audio. In the In-Distribution DDPM settings, the environmental sound trigger shows an SNR of 30.292, followed by Infrasound at 28.878, both significantly better than other watermark trigger options. For In-Distribution DDIM settings, the environmental sound trigger leads in SNR at 28.987, followed by the patch at 26.412 and Infrasound at 26.303. These results validate our hypothesis that invisible triggers, such as Infrasound and environmental sounds, have a minimal impact on the generated audio watermark while maintaining relatively high WSR and F1 scores when compared to random noise-based triggers and special triggers like the patch and Hello Kitty.

\begin{figure}[t]
    \centering
    \begin{minipage}[b]{0.48\linewidth}
        \includegraphics[width=1.05\linewidth]{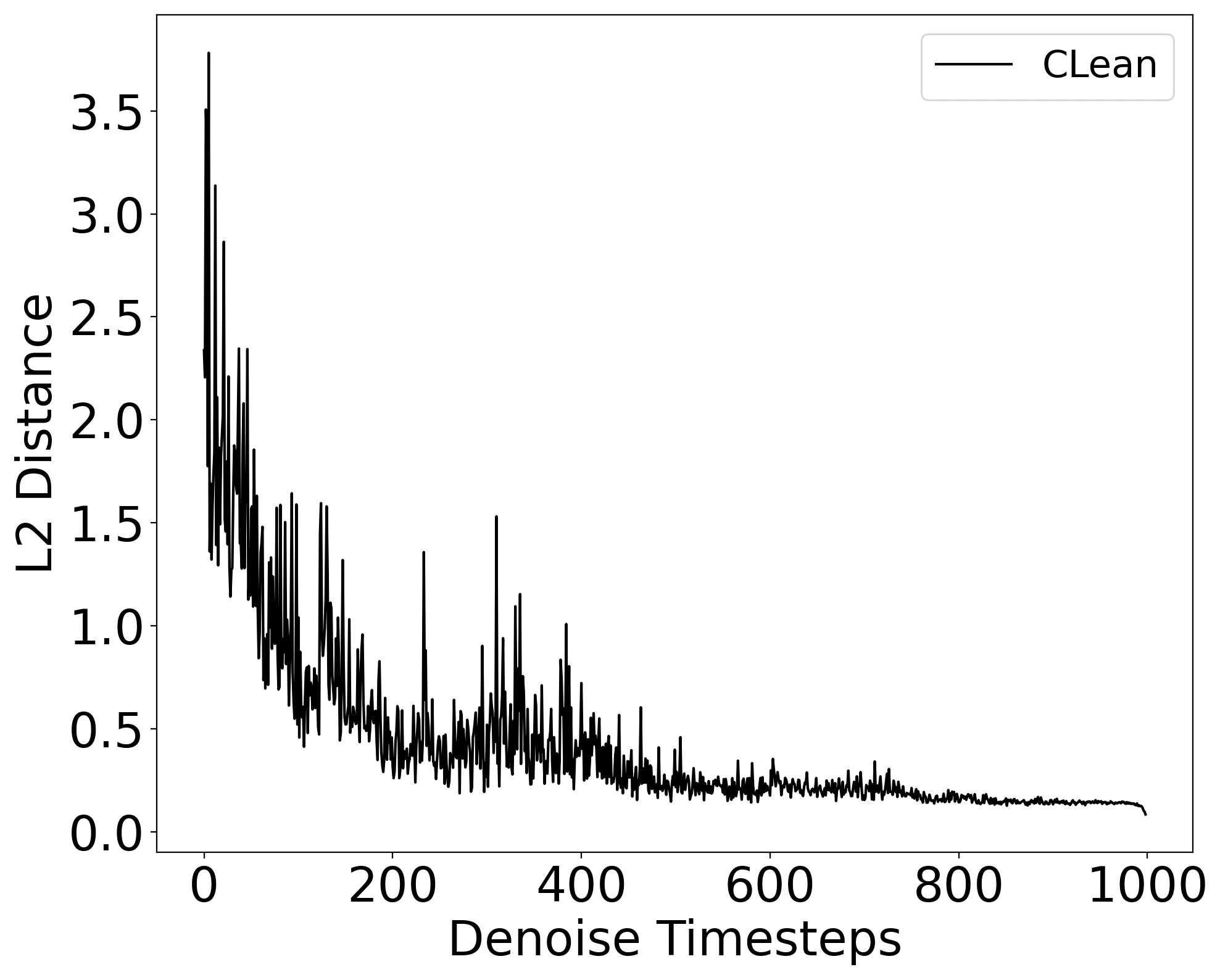}
        \begin{subcaption}{\small Vanilla Model}
            \label{fig:image1}
        \end{subcaption}
    \end{minipage}
    \hfill
    \begin{minipage}[b]{0.48\linewidth}
        \includegraphics[width=1.05\linewidth]{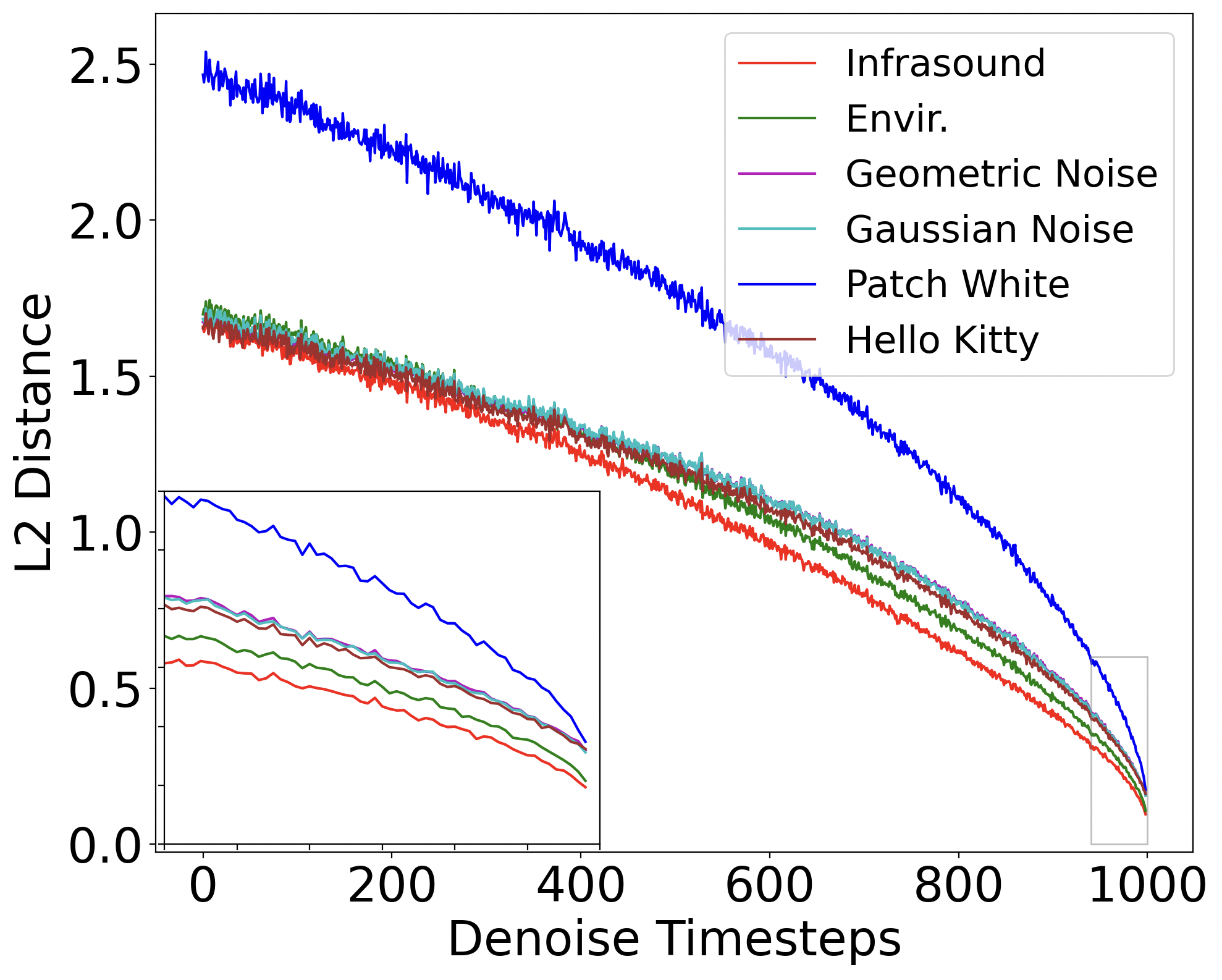}
        \begin{subcaption}{\small Watermarking Model}
            \label{fig:image2}
        \end{subcaption}
    \end{minipage}
    \caption{L2 distance for vanilla model and watermarking model in the denoising process with DDPM sampling method. Infrasound demonstrates the fastest denoise convergence speed followed by the environmental sound. We also zoom in from 940 to 1000 steps to see the detail in the graph.}
\label{fig:l2dis}
\vspace{-0.2cm}
\end{figure}
\setcounter{subplot}{0}  

\textbf{Remarks.} In \textbf{Figure~\ref{fig:cor}}, we can observe that there is a trade-off between FID and WSR with respect to different image-level triggers, namely the Patch and Hello Kitty. Specifically, When Patch White is employed as a watermark trigger in the In-Distribution DDIM setting, it delivers the best performance in terms of both FID and IS, albeit at the cost of achieving the lowest WSR. A possible explanation is that the small white patch can provide only insufficient information because it only takes up a few pixels at the image level, and therefore less alteration for the feature vector level. As the result shows, when using Patch White as the watermark trigger, it has almost no impact on the model utility and less distortion of the generated samples. However, having a small patch with size of 3 leads to inaccurate watermark generation, and it hurts the watermarking model by deviating the model's attention away from itself. Conversely, when using Hello Kitty as the watermark trigger in In-Distribution DDPM, we can see that it achieves the highest WSR while suffering from a relatively significant degradation in terms of FID and IS. It is possible that the conspicuous characteristics of Hello Kitty lead to a better watermark generation but at the expense of model utility. By comparison, our proposed invisible watermark triggers, e.g. Infrasound, Envir., and even Geometric noise and Gaussian noise, exhibit robustness across all evaluation metrics. Specifically, our invisible watermark triggers generally achieve better performance compared to Hello Kitty, but inferior to the patch trigger in terms of FID.

Furthermore, as illustrated in \textbf{Figure~\ref{fig:l2dis}}, our proposed invisible watermark triggers can converge faster than image-level triggers. Among the six watermarking triggers, Infrasound stands out as the fastest denoise convergence speed followed by environmental sound. The two invisible watermark triggers are also accompanied by better generation quality in terms of closer L2 distance to the watermark target (see the zoom-in part of Figure~\ref{fig:image2}). A similar observation is illustrated in Figure~\ref{fig:denoise}. This trend demonstrates the benefit of those watermark triggers that are invisible at the mel-spectrogram level and imperceptible at the audio level during the watermarking process.

\begin{figure*}[t]
  \centering
  \includegraphics[width=\textwidth]{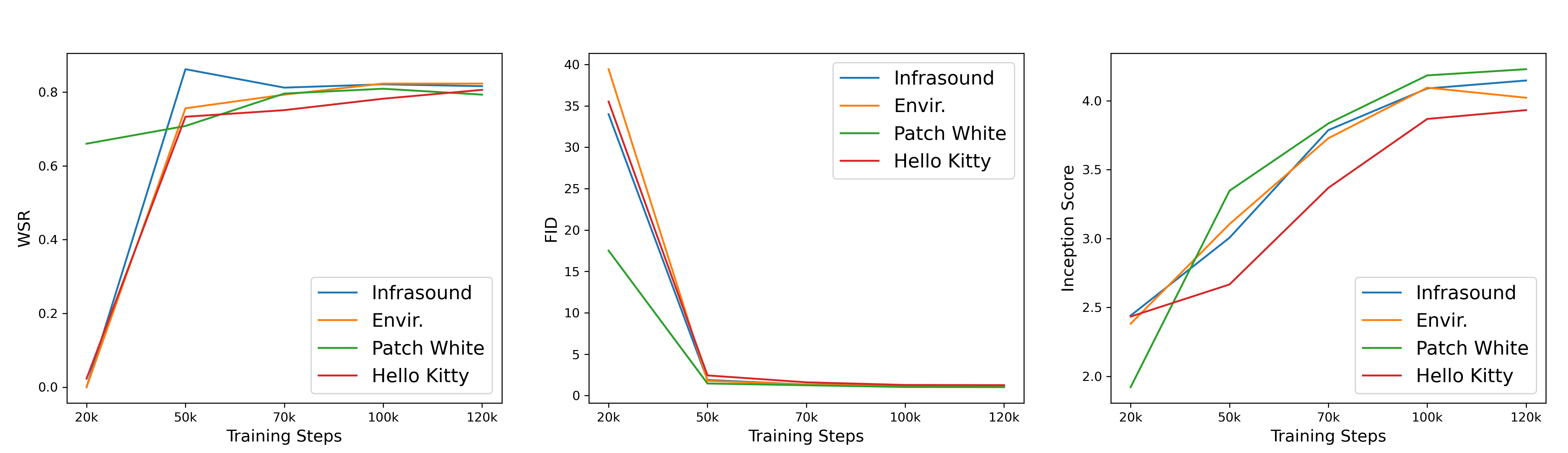}
  \caption{The relationship between training steps and effectiveness of the DDIM model, measured by WSR $(\uparrow)$, FID $(\downarrow)$, IS $(\uparrow)$.}
\label{fig:ablation_train_steps}
\end{figure*}

\subsection{Ablation Studies} 
\label{sec:ablation}

\textbf{Training steps}. We first study the impact of the training steps on the performance of audio diffusion model watermarking. We examine the watermarking performance of In-Distribution DDIM at different training steps. Note that this training step is different from the denoising time step as in Figure~\ref{fig:l2dis}. The training step here refers to the iterations during training, while the denoising time step refers to the reverse sampling process during inference. In \textbf{Figure~\ref{fig:ablation_train_steps}}, we present the results obtained from audio diffusion models trained with Infrasound, Envir., Patch White, and Hello Kitty as watermark triggers. We make three observations. (1) when the training step is limited, e.g., 20k, most cases reach a WSR of 0\%, with the exception of Patch White. (2) After 50k training steps, the watermarking models with different triggers can achieve more than 70\% WSR, indicating that these watermark triggers can be easily implanted into the diffusion models. As training proceeds, the watermark performance of the models could be enhanced and eventually the training converges around 100k steps. (3) Among the four watermark triggers, Infrasound and environmental sound demonstrate faster training convergence, indicating that they are more efficient to be inserted into the audio diffusion models when compared to the patch and hello kitty trigger. The faster training convergence allows the two invisible triggers to achieve higher WSR, lower FID, and higher IS in earlier training steps. However, as previously shown, the superior generation quality of the patch trigger comes at the cost of lower WSR, highlighting a trade-off between generation quality and watermarking robustness.

\begin{figure*}
  \centering
  \includegraphics[width=0.9\textwidth]{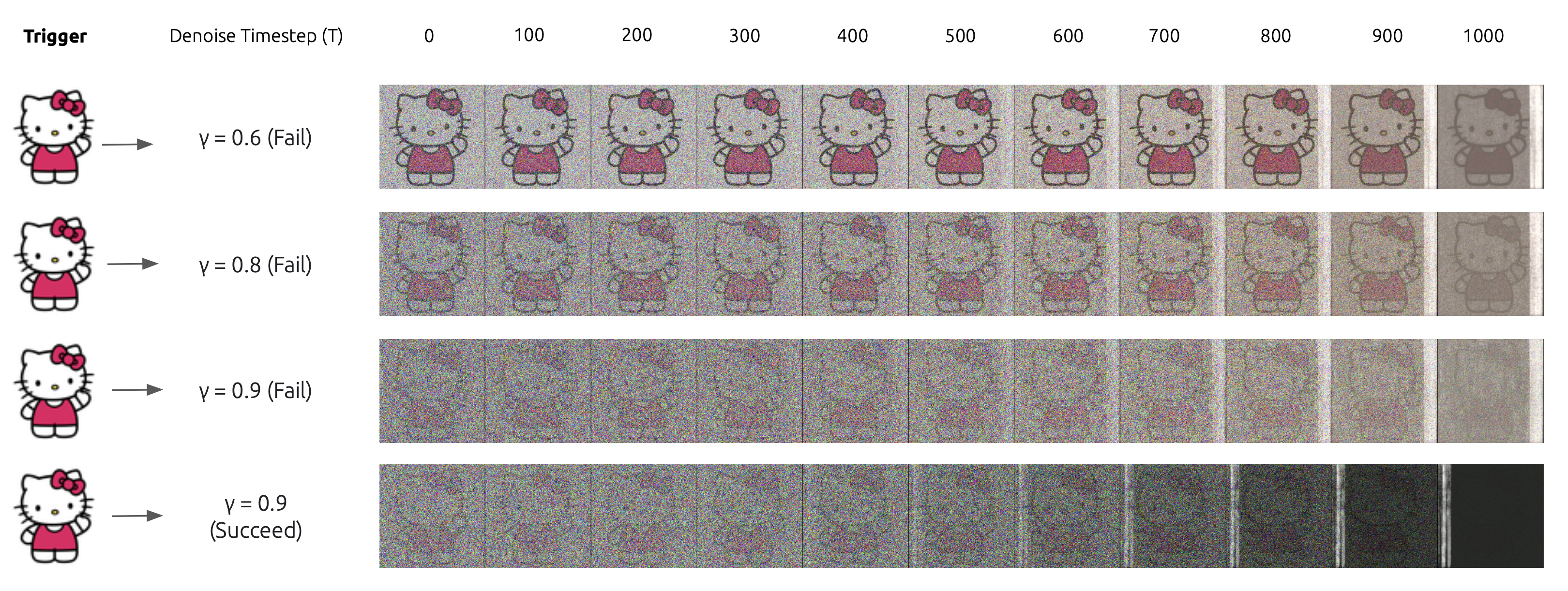}
  \caption{Audio diffusion watermarking process under
256×256 resolutions of Mel-Spectrogram. }
\label{fig:6}
\end{figure*}

\textbf{Instance-Specific Watermark}. Following prior works~\cite{zhao2023recipe,liu2023watermarking}, we further study watermarking audio diffusion models using a specific instance as the watermark. In the experiment, we select the spoken word ``wow" from the Speech Command dataset~\cite{warden2018speech} as the target watermark and employ Infrasound and Hello Kitty as watermark triggers to train the diffusion models, respectively. In this case, if the input noise is mixed with the watermark trigger efficiently, an effective watermarking model should generate the identical mel-spectrogram to that of the target watermark instance. This also enables the pairwise comparison of the generated audio and the exact spoken word ``wow", using SSIM~\cite{1284395} and PSNR~\cite{hore2010image} for mel-spectrogram quality, as well as SNR for audio quality. The Structure Similarity Index Measure (SSIM) is a widely used image quality assessment metric that quantifies the structural similarity between two images by taking into account luminance, contrast, and structure. In our case, the ``image" refers to the mel-spectrograms of the target instance watermark and the generated watermark. The Peak Signal-to-Noise (PSNR) measures the ratio of the maximum possible signal strength to the noise introduced by the compression or distortion process. A higher SSIM or PSNR value indicates that the reconstructed or compressed signal is closer in quality to the original reference signal and has less distortion or noise. 

\begin{table}[t]
\centering
\scalebox{0.97}{
\small  
\begin{tabular}{|c|c|c|c|c|}
 \hline
&  PSNR $(\uparrow)$ & SSIM $(\uparrow)$ & MSE $(\downarrow)$ & SNR $(\uparrow)$ \\
 \hline
Infrasound  & 97.615  &  0.966    & \(1.45 \times 10^{-5}\) & -0.038 \\ \hline
Hello Kitty & 97.124  &  0.960    &  \(1.7 \times 10^{-5}\) &  -0.04  \\
 \hline
\end{tabular}
}
\caption{DDPM Performance of instance-specific audio diffusion watermarking, which has 100\% watermarking success rate. }
\label{table:d2i_examples}
\end{table}

\textbf{Table~\ref{table:d2i_examples}} shows the results and we make two observations. (1) The samples generated from both watermark triggers can achieve a high SSIM and PSNR, indicating high watermarking success rate in the instance-specific watermark setting. Similar to the In-Distribution and Out-of-Distribution settings, the watermark trigger Infrasound offers slightly better results than the Hello Kitty regarding the watermark quality: higher PSNR, SSIM and lower MSE. (2) The SNR evaluates the ratio of the strength of a useful signal to the strength of background noise that can potentially degrade the quality of the signal. A higher SNR value indicates that the signal is stronger in relation to the noise, which generally results in better signal quality. In the last column of Table~\ref{table:d2i_examples}, we can also see similar result pattern for both of the watermark triggers, they are all below 1 for SNR score, indicating that the noise is very huge. A possible explanation is that the noise addition and denoising process typically will bring noise to the data distribution and these injected noises are often human inaudible or invisible but machine capture-able. 




\begin{table}[t]
\centering
\scalebox{0.85}{
\small  
\renewcommand{\arraystretch}{1}  
\begin{tabular}{|c|cc|cc|}
 \hline
 & \multicolumn{2}{c|}{SNR} & \multicolumn{2}{c|}{WSR} \\
 \hline
  Trigger & Hello Kitty & Infrasound & Hello Kitty & Infrasound \\
 \hline
 $\gamma$ = 0.6            & 8.65      & 7.02   & 0.009    & 0.011 \\
 $\gamma$ = 0.8            & 8.95      & 8.84   & 0.006    & 0.010 \\
 $\gamma$ = 0.9 (failed)      & 6.16      & 9.77   & 0.013    & 0.027 \\
 $\gamma$ = 0.9 (succeed)  & 19.23     & 21.01  & --       & --  \\ 
 \hline
\end{tabular}
}
\caption{DDPM performance of audio diffusion watermarking under 256$\times$256 resolutions of Mel-Spectrogram. For the last row, we measure the SNR for successful watermarking instances. }
\label{table:resolution}
\end{table}

\textbf{Resolution of Mel-Spectrogram.} In this set of experiment, we study the effect of the resolution of the mel-spectrograms in watermarking audio diffusion models. The higher the resolution of mel-spectrograms, the better the quality of the resulting audio after conversion. Therefore, we set the resolution of the mel-spectrograms to 256$\times$256. Due to limited computing resources, we set the batch size to 2 and 
train the model using Infrasound and Hello Kitty as watermark triggers for comparison. As illustrated in \textbf{Figure~\ref{fig:6}}, we find that the model fails to denoise the watermark trigger in many cases. This results in the low WSR for 120k training steps as shown in \textbf{Table~\ref{table:resolution}}. 

We next look at the blending ratio $\gamma$ of the watermark trigger to see if they can improve the performance. We set $\gamma$ to $\{0.6, 0.8, 0.9\}$ for training. The higher the value of $\gamma$, the more transparent the watermark trigger. As $\gamma$ increases, we find that the watermarking model still fails to denoise the watermark trigger in most cases, but with a slightly better performance. In the meantime, we make similar observations that invisible watermark triggers like Infrasound show superior performance when compared to the Hello Kitty, both in terms of better generation quality with a higher SNR and a more robust watermarking process with a higher WSR.

\section{Discussion}
\label{sec:discussion}

As mentioned in Section~\ref{sec:4.2}, our watermarking diffusion model adopts a strategy similar to backdoor attacks on diffusion models~\cite{chen2023trojdiff}, but is tailored for safeguarding generative audio diffusion models. The experimental outcomes provide valuable insights, which align with prior research on backdoor attacks within the context of diffusion models: (1) Compared to the vanilla diffusion model, the FID scores of our watermarking diffusion models do not suffer from degradation. In fact, the watermarking diffusion models often yield improved performance. This observation mirrors findings in Baddiffusion~\cite{chou2023backdoor}, where the introduction of a backdoor trigger led to a significant enhancement in FID scores. (2) Baddiffusion~\cite{chou2023backdoor} studies and verifies the applicability of their approach to high-resolution dataset~\cite{liu2015deep}. Specifically, they employed a batch size of 64 and fine-tuned the diffusion model for 100 epochs. However, as discussed in Section~\ref{sec:ablation}, our attempts to train diffusion models from scratch failed to effectively denoise the watermark trigger with a batch size of 2. 
While we cannot conduct experiments for 256$\times$256 with a batch size larger than 2, 
\cite{smith2017don} shows that increasing batch size could improve the quality of generated samples, as well as increasing the training time. Therefore, our observation suggests that if using higher resolution, a large $\gamma$, batch size and training step may be necessary to facilitate successful model convergence and watermark injection. Unluckily, these adjustment may also alternatively enable the backdoor attacks for higher-resolution diffusion generation. 

In addition, the different denoising process schedules between the vanilla model and the watermarking model, as illustrated in Figure~\ref{fig:l2dis}, indicate a potential defense strategy for detecting backdoored diffusion models through the observation of denoising discrepancies. This akins to the different model learning capabilities for the clean and the backdoor portions of data observed in deep learning models~\cite{li2021anti}.

\section{Conclusion}
\label{sec:conclusion}
In this paper, we have presented the first watermarking method for audio diffusion models. We next demonstrated the pivotal role of selecting the appropriate watermark trigger in watermarking audio diffusion models, featuring two invisible watermark triggers, Infrasound and environmental sound. 
These watermark triggers options benefit from invisible to the model-stealing attackers both at the audio level and the mel-spetrogram level, preventing the duplication of watermarks for intellectual property violation. 
Extensive experiments demonstrate that the two invisible watermark triggers achieve high watermarking success rate and low FID when compared to existing common watermarking triggers. We believe this work opens new avenues for robust model protection and data copyright in audio-based diffusion models and generative AI research.

\bibliographystyle{IEEEtran}                                    
\bibliography{output.bbl}

\begin{thebibliography}{10}
\providecommand{\url}[1]{#1}
\csname url@samestyle\endcsname
\providecommand{\newblock}{\relax}
\providecommand{\bibinfo}[2]{#2}
\providecommand{\BIBentrySTDinterwordspacing}{\spaceskip=0pt\relax}
\providecommand{\BIBentryALTinterwordstretchfactor}{4}
\providecommand{\BIBentryALTinterwordspacing}{\spaceskip=\fontdimen2\font plus
\BIBentryALTinterwordstretchfactor\fontdimen3\font minus
  \fontdimen4\font\relax}
\providecommand{\BIBforeignlanguage}[2]{{%
\expandafter\ifx\csname l@#1\endcsname\relax
\typeout{** WARNING: IEEEtran.bst: No hyphenation pattern has been}%
\typeout{** loaded for the language `#1'. Using the pattern for}%
\typeout{** the default language instead.}%
\else
\language=\csname l@#1\endcsname
\fi
#2}}
\providecommand{\BIBdecl}{\relax}
\BIBdecl

\bibitem{brock2018large}
A.~Brock, J.~Donahue, and K.~Simonyan, ``Large scale gan training for high
  fidelity natural image synthesis,'' \emph{arXiv preprint arXiv:1809.11096},
  2018.

\bibitem{kingma2013auto}
D.~P. Kingma and M.~Welling, ``Auto-encoding variational bayes,'' \emph{arXiv
  preprint arXiv:1312.6114}, 2013.

\bibitem{rombach2022high}
R.~Rombach, A.~Blattmann, D.~Lorenz, P.~Esser, and B.~Ommer, ``High-resolution
  image synthesis with latent diffusion models,'' in \emph{Proceedings of the
  IEEE/CVF conference on computer vision and pattern recognition}, 2022, pp.
  10\,684--10\,695.

\bibitem{ramesh2022hierarchical}
A.~Ramesh, P.~Dhariwal, A.~Nichol, C.~Chu, and M.~Chen, ``Hierarchical
  text-conditional image generation with clip latents,'' \emph{arXiv preprint
  arXiv:2204.06125}, vol.~1, no.~2, p.~3, 2022.

\bibitem{liu2023audioldm}
H.~Liu, Z.~Chen, Y.~Yuan, X.~Mei, X.~Liu, D.~Mandic, W.~Wang, and M.~D.
  Plumbley, ``Audioldm: Text-to-audio generation with latent diffusion
  models,'' \emph{arXiv preprint arXiv:2301.12503}, 2023.

\bibitem{verdoliva2020media}
L.~Verdoliva, ``Media forensics and deepfakes: an overview,'' \emph{IEEE
  Journal of Selected Topics in Signal Processing}, vol.~14, no.~5, pp.
  910--932, 2020.

\bibitem{yu2021artificial}
N.~Yu, V.~Skripniuk, S.~Abdelnabi, and M.~Fritz, ``Artificial fingerprinting
  for generative models: Rooting deepfake attribution in training data,'' in
  \emph{Proceedings of the IEEE/CVF International conference on computer
  vision}, 2021, pp. 14\,448--14\,457.

\bibitem{zhao2023recipe}
Y.~Zhao, T.~Pang, C.~Du, X.~Yang, N.-M. Cheung, and M.~Lin, ``A recipe for
  watermarking diffusion models,'' \emph{arXiv preprint arXiv:2303.10137},
  2023.

\bibitem{wen2023tree}
Y.~Wen, J.~Kirchenbauer, J.~Geiping, and T.~Goldstein, ``Tree-ring watermarks:
  Fingerprints for diffusion images that are invisible and robust,''
  \emph{arXiv preprint arXiv:2305.20030}, 2023.

\bibitem{liu2023watermarking}
Y.~Liu, Z.~Li, M.~Backes, Y.~Shen, and Y.~Zhang, ``Watermarking diffusion
  model,'' \emph{arXiv preprint arXiv:2305.12502}, 2023.

\bibitem{ho2020denoising}
J.~Ho, A.~Jain, and P.~Abbeel, ``Denoising diffusion probabilistic models,''
  \emph{Advances in neural information processing systems}, vol.~33, pp.
  6840--6851, 2020.

\bibitem{song2020denoising}
J.~Song, C.~Meng, and S.~Ermon, ``Denoising diffusion implicit models,''
  \emph{arXiv preprint arXiv:2010.02502}, 2020.

\bibitem{dhariwal2021diffusion}
P.~Dhariwal and A.~Nichol, ``Diffusion models beat gans on image synthesis,''
  \emph{Advances in neural information processing systems}, vol.~34, pp.
  8780--8794, 2021.

\bibitem{kong2021diffwave}
Z.~Kong, W.~Ping, J.~Huang, K.~Zhao, and B.~Catanzaro, ``Diffwave: A versatile
  diffusion model for audio synthesis,'' in \emph{International Conference on
  Learning Representations}, 2021.

\bibitem{song2019generative}
Y.~Song and S.~Ermon, ``Generative modeling by estimating gradients of the data
  distribution,'' \emph{Advances in neural information processing systems},
  vol.~32, 2019.

\bibitem{song2020score}
Y.~Song, J.~Sohl-Dickstein, D.~P. Kingma, A.~Kumar, S.~Ermon, and B.~Poole,
  ``Score-based generative modeling through stochastic differential
  equations,'' \emph{arXiv preprint arXiv:2011.13456}, 2020.

\bibitem{ruanaidh1996watermarking}
J.~{\'O}. Ruanaidh, W.~Dowling, and F.~Boland, ``Watermarking digital images
  for copyright protection,'' \emph{IEE Proceedings-Vision, Image and Signal
  Processing}, vol. 143, no.~4, pp. 250--256, 1996.

\bibitem{cox1996secure}
I.~J. Cox, J.~Kilian, T.~Leighton, and T.~Shamoon, ``Secure spread spectrum
  watermarking for images, audio and video,'' in \emph{Proceedings of 3rd IEEE
  international conference on image processing}, vol.~3.\hskip 1em plus 0.5em
  minus 0.4em\relax IEEE, 1996, pp. 243--246.

\bibitem{ong2021protecting}
D.~S. Ong, C.~S. Chan, K.~W. Ng, L.~Fan, and Q.~Yang, ``Protecting intellectual
  property of generative adversarial networks from ambiguity attacks,'' in
  \emph{Proceedings of the IEEE/CVF Conference on Computer Vision and Pattern
  Recognition}, 2021, pp. 3630--3639.

\bibitem{fei2022supervised}
J.~Fei, Z.~Xia, B.~Tondi, and M.~Barni, ``Supervised gan watermarking for
  intellectual property protection,'' in \emph{2022 IEEE International Workshop
  on Information Forensics and Security (WIFS)}.\hskip 1em plus 0.5em minus
  0.4em\relax IEEE, 2022, pp. 1--6.

\bibitem{chen2023trojdiff}
W.~Chen, D.~Song, and B.~Li, ``Trojdiff: Trojan attacks on diffusion models
  with diverse targets,'' in \emph{Proceedings of the IEEE/CVF Conference on
  Computer Vision and Pattern Recognition}, 2023, pp. 4035--4044.

\bibitem{chou2023backdoor}
S.-Y. Chou, P.-Y. Chen, and T.-Y. Ho, ``How to backdoor diffusion models?'' in
  \emph{Proceedings of the IEEE/CVF Conference on Computer Vision and Pattern
  Recognition}, 2023, pp. 4015--4024.

\bibitem{warden2018speech}
P.~Warden, ``Speech commands: A dataset for limited-vocabulary speech
  recognition,'' \emph{arXiv preprint arXiv:1804.03209}, 2018.

\bibitem{piczak2015esc}
K.~J. Piczak, ``Esc: Dataset for environmental sound classification,'' in
  \emph{Proceedings of the 23rd ACM international conference on Multimedia},
  2015, pp. 1015--1018.

\bibitem{wei2020framework}
W.~Wei, L.~Liu, M.~Loper, K.-H. Chow, M.~E. Gursoy, S.~Truex, and Y.~Wu, ``A
  framework for evaluating client privacy leakages in federated learning,'' in
  \emph{Proc. Eur. Symp. Res. Comput. Secur. (ESORICS)}.\hskip 1em plus 0.5em
  minus 0.4em\relax Springer, 2020, pp. 545--566.

\bibitem{xie2017aggregated}
S.~Xie, R.~Girshick, P.~Doll{\'a}r, Z.~Tu, and K.~He, ``Aggregated residual
  transformations for deep neural networks,'' in \emph{Proceedings of the IEEE
  conference on computer vision and pattern recognition}, 2017, pp. 1492--1500.

\bibitem{Szegedy_2016_CVPR}
C.~Szegedy, V.~Vanhoucke, S.~Ioffe, J.~Shlens, and Z.~Wojna, ``Rethinking the
  inception architecture for computer vision,'' in \emph{Proceedings of the
  IEEE Conference on Computer Vision and Pattern Recognition (CVPR)}, June
  2016.

\bibitem{heusel2017gans}
M.~Heusel, H.~Ramsauer, T.~Unterthiner, B.~Nessler, and S.~Hochreiter, ``Gans
  trained by a two time-scale update rule converge to a local nash
  equilibrium,'' \emph{Advances in neural information processing systems},
  vol.~30, 2017.

\bibitem{salimans2016improved}
T.~Salimans, I.~Goodfellow, W.~Zaremba, V.~Cheung, A.~Radford, and X.~Chen,
  ``Improved techniques for training gans,'' \emph{Advances in neural
  information processing systems}, vol.~29, 2016.

\bibitem{kynkaanniemi2019improved}
T.~Kynk{\"a}{\"a}nniemi, T.~Karras, S.~Laine, J.~Lehtinen, and T.~Aila,
  ``Improved precision and recall metric for assessing generative models,''
  \emph{Advances in Neural Information Processing Systems}, vol.~32, 2019.

\bibitem{1284395}
Z.~Wang, A.~Bovik, H.~Sheikh, and E.~Simoncelli, ``Image quality assessment:
  from error visibility to structural similarity,'' \emph{IEEE Transactions on
  Image Processing}, vol.~13, no.~4, pp. 600--612, 2004.

\bibitem{hore2010image}
A.~Hore and D.~Ziou, ``Image quality metrics: Psnr vs. ssim,'' in \emph{2010
  20th international conference on pattern recognition}.\hskip 1em plus 0.5em
  minus 0.4em\relax IEEE, 2010, pp. 2366--2369.

\bibitem{liu2015deep}
Z.~Liu, P.~Luo, X.~Wang, and X.~Tang, ``Deep learning face attributes in the
  wild,'' in \emph{Proceedings of the IEEE international conference on computer
  vision}, 2015, pp. 3730--3738.

\bibitem{smith2017don}
S.~L. Smith, P.-J. Kindermans, C.~Ying, and Q.~V. Le, ``Don't decay the
  learning rate, increase the batch size,'' \emph{arXiv preprint
  arXiv:1711.00489}, 2017.

\bibitem{li2021anti}
Y.~Li, X.~Lyu, N.~Koren, L.~Lyu, B.~Li, and X.~Ma, ``Anti-backdoor learning:
  Training clean models on poisoned data,'' \emph{Advances in Neural
  Information Processing Systems}, vol.~34, pp. 14\,900--14\,912, 2021.

\end{thebibliography}

\end{document}